\documentclass[reprint, amssymb, amsmath, aip,cha]{revtex4-1}

\usepackage{bm}%
\usepackage{graphicx,color}
\usepackage{hyperref}%
\expandafter\ifx\csname package@font\endcsname\relax\else
 \expandafter\expandafter
 \expandafter\usepackage
 \expandafter\expandafter
 \expandafter{\csname package@font\endcsname}%
\fi
\begin{document}

\title{Rates of transesterification in epoxy-thiol vitrimers  }
\author{Alexandra Gablier, Mohand O. Saed and Eugene M. Terentjev}
\affiliation{Cavendish Laboratory, University of Cambridge,  J.J. Thomson Avenue, Cambridge, CB3 0HE, U.K.}

\date{\today}

\begin{abstract}
\noindent Vitrimers, an important subset of dynamically crosslinked polymer networks, have many technological applications for their excellent properties, and the ability to be re-processed through plastic flow above the so-called vitrification temperature. We report a simple and efficient method of generating such adaptive crosslinked networks relying on transesterification for their bond exchange by utilising the `click' chemistry of epoxy and thiols, which also has the advantage of a low glass transition temperature. We vary the chemical structure of thiol spacers to probe the effects of concentration and the local environment of ester groups on the macroscopic elastic-plastic transition. The thermal activation energy of transesterification bond exchange is determined for each chemical structure, and for a varying concentration of catalyst, establishing the conditions for the optimal, and for the suppressed bond exchange. However, we also discover that the temperature of elastic-plastic transition is strongly affected by the stiffness (dynamic rubber modulus) of the network, with softer networks having a much lower vitrification temperature even when their bond-exchange activation energy is higher. This combination of chemical and physical control factors should help optimise the processability of vitrimer plastics
\end{abstract}

%\pacs{eventually need to fix this -- 82.35.Gh, 05.70.Np}

\maketitle

%%%%%%%%%%%%%%%%%%%%%%%%%%%%%%%%%%%%%%%%%%
\section{Introduction}
Vitrimers are covalent polymer networks capable of associative thermally-activated bond exchange reactions (BER), a concept originally formulated in 2011 by Leibler et al.\cite{leibler2011} Due to the temperature dependence of the exchange reaction rate, such networks become dynamic at elevated temperatures, enabling network re-configuration through the reshuffling of covalent bonds (while keeping the total number of bonds constant at all times). Upon cooling the material, the reaction rate decreases and the network structure becomes stable (“frozen”) beyond a certain point, which Leibler and others defined as the topology freezing temperature, Tv. Below it, the bond-exchange rate is negligible, and the material behaves as a standard thermoset.\cite{denissen2016,krishna2020} Since vitrimers are malleable at high temperatures, they have the capability to self-heal,\cite{cash2015} be remoulded \cite{chen2019} and  reprogrammed \cite{xlce2014,yanji2018} in a manner reminiscent of thermoplastics, while maintaining the properties of thermosets at lower operation temperatures. 

Reactions having been demonstrated as effective for associative bond-exchange in vitrimers include boron-based exchanges,\cite{leibler2017bo,gablier2019,guan2015} olefin metathesis,\cite{guan2012} and transamination of vinylogous urethanes \cite{denissen2015,denissen2017,guerre2018,duprez2017} among others.\cite{chen2019,guan2017,fortman2015,ishibashi2018,hendriks2017,obadia2015,tang2017} Despite this diversity, the transesterification reaction remains the main focus in terms of applications involving vitrimers,\cite{leibler2012cat,leibler2012met,yanji2014,yanji2016,zhao2019,Hillmyer2014,chen2017,hayashi2019a,epoxy2019,shi2017,catalyst-free2018} due as much to its history as the first reaction used for vitrimers, as to the ease of developing vitrimers with such an exchange mechanism. Vitrimers based on the transesterification BER are most commonly formed through an epoxy-acid polymerisation; the attractiveness of this chemistry is due to the plethora of available starting materials. However, random branching occurs due to an uncontrolled side reaction (e.g. etherification, condensation esterification, and disproportionation) between the
carboxylic acids and the hydroxyl groups arising from the ring opening of the epoxy (esterification).\cite{matejka1982}
 The resulting crosslinking density, as well as the network homogeneity, are therefore uncontrollable in vitrimers produced by epoxy-acid reaction. Additionally, epoxy-acid polymerisation requires high temperatures and extended periods of time (up to 180$^\circ$C for multiple hours).\cite{xlce2014} Here we use an alternative way of forming transesterification-based vitrimers, based on epoxy-thiol chemistry,\cite{li2018} which addresses these shortcomings.

The current theoretical understanding of vitrimer materials rests on the characteristic activation energy and the rate of attempts of the BER, which in turn determine the elastic-plastic transition kinetics, and the network flow properties.\cite{leibler2011,meng2016,meng2019} Experimentally, the nature and the concentration of catalyst were demonstrated to modulate the elastic-to-plastic transition.\cite{leibler2012cat,self2018} Initial studies were conducted on the role of hydroxyl functions in the network,\cite{snyder2018} the influence of the glass transition of the material,\cite{yu2014} or the impact of network crosslink density on the BER rate,\cite{zhao2019,hayashi2019b,brutman2019} in a range of different vitrimers. However, the existing studies have several coupled parameters involved, so it remains hard to draw clear conclusions. These restrictions stem from the limited control of the network structure in the systems used so far. Comprehensive studies of vitrimer networks based on transesterification, aiming at cleanly isolating and quantifying the factors influencing the vitrimer properties on the macroscopic scale are lacking. 

Here, stable transesterification-based vitrimer networks with a controlled topology are obtained via an epoxy-thiol polymerisation. By varying the network composition, we investigate the influence of such factors as the catalyst concentration, the concentration of reactive bonds within the network, the activation energy of transesterification, and the rubber modulus of the network, on the elastic-plastic transition of the vitrimers. We find that two key factors determine the kinetics of this transition: the elastic stiffness of the network, and the concentration of exchangeable bonds. 

\begin{figure} 
\centering
\includegraphics[width=\columnwidth]{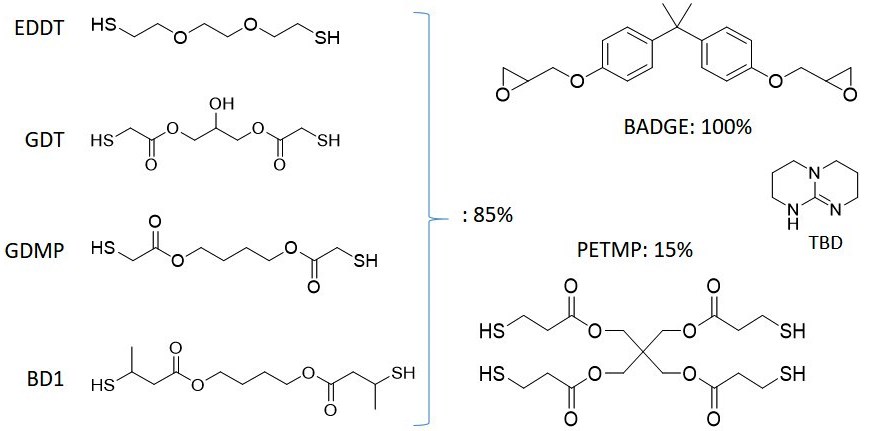}
\caption{An epoxy monomer BADGE was reacted with a di-thiol spacer (from the list of four) and the 4-functional thiol crosslinker (PETMP) in the presence of TBD catalyst.  A crosslinking density of 15 mol\% was used unless specified otherwise; it is defined as the proportion of thiol functions belonging to the crosslinker with respect to the total thiol functions – which are in the 1:1 stoichiometric ratio to epoxy functions.}
\label{scheme1}
\end{figure}

Our synthesis is based on the reaction of thiol and epoxy, an example of `click' chemistry that has recently been increasingly popular.\cite{bowman2010,brandle2012,nicolay2014}. Ultimately the networks built by this reaction allow the same transesterification exchange, \cite{li2018,belmonte2017,jin2016,fernandez2016}  but resulting in robust and homogeneous networks compared to the traditional epoxy-acid chemistry.\cite{khan2016,konuray2017} Compared to an epoxy-acid polymerisation, the thiol-based chemistry has a minimum of side reactions, resulting in full control over the crosslinking density of the network and the overall network structure after polymerization. In selecting our starting materials, we deliberately stay close to the original work of Leibler,\cite{leibler2011,leibler2012cat} using the same epoxy monomer and catalyst for the bond exchange, while testing a variety of thiol spacers with different environments for the ester and hydroxyl groups (see Fig. \ref{scheme1}).

\begin{figure} 
\centering
\includegraphics[width=0.8\columnwidth]{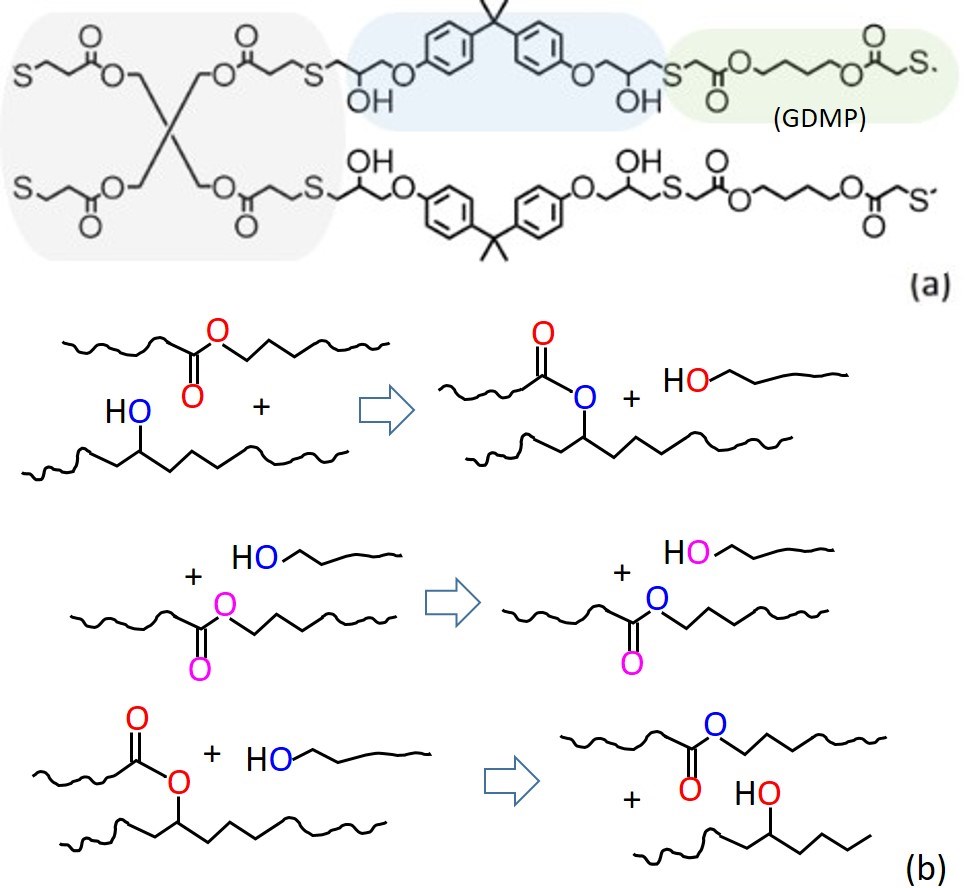}
\caption{(a) The network structure showing the ester groups arising from the crosslinker and (potentially) the thiol spacers, and the hydroxyl groups arising from the reacted epoxy (and the GDT spacer, as an example). The three elements of the network are shaded to highlight their correspondence to the initial monomers. (b) The illustration of transesterification re-connection of polymer chains in the network, with matching oxygens coloured for easier tracking. Two chains could re-connect into a 3-functional junction and a dangling end, the dangling end with a chain preserves the topology on re-connection, and the 3-functional junction with a dangling end re-connect into the two chains. }
\label{scheme2}
\end{figure}

The illustration in Fig. \ref{scheme2}(a) shows the controlled network topology, where no spontaneous random crosslinks can occur. Figure \ref{scheme2}(b) illustrates the reconnection of polymer strands in the key pathways of transesterification exchange between the ester and the hydroxyl groups. At the first stage, the initially linear chains transform into a three-functional junction and a free dangling end rich in hydroxyl groups. Subsequently, this free dangling end could exchange with a linear chain segment, which retains the same topology even though the local chain tension would be released. Alternatively, this free end may exchange with an ester group of the triple-junction, which the recovers the topology of two linear chains. After a large number of such transesterification exchanges, the network would adopt a statistical equilibrium configuration with some fraction of additional triple junctions, and some fraction of free dangling ends acting as a network plasticiser. 

%%%%%%%%%%%%%%%%%%%%%%%%%%%%%%%%%%%%%%%%%%%%%%
\section{Experimental section}
\textbf{Materials}. All reagents were purchased from commercial suppliers and used as received. 1,5,7-triazabicyclo[4.4.0]dec-5-ene (TBD), pentaerythritol tetrakis (3-mercaptopropionate) (PETMP), 2,2’-(ethylenedioxy)diethanethiol (EDDT), and bisphenol A diglycidyl ether (BADGE) were purchased from Sigma-Aldrich. 1,4-Bis (3-mercaptobutyryloxy)butane (BD1) was provided by Showa Denko; 1,4-butanediol bis(thioglycolate) (GDMP) was sourced from TCI, and glyceryl dithioglycolate (GDT) was obtained from Bruno Bock Chemische Fabrik GmbH \& Co. TBD was stored in a desiccator, because its catalytic characteristics are very strongly affected by humidity. All polymerisation reactions were performed in dry solvents. Anhydrous acetonitrile was purchased from Sigma-Aldrich and used as such; toluene was purchased from Fischer and dried over 3\AA activated molecular sieves. Purity of reagents was accounted for when calculating experimental masses for all monomers for polymerisations. 

\textbf{Preparation of polymer networks}. 
Our networks were polymerised from a di-functional epoxy monomer (BADGE), with a di-functional thiol spacer (EDDT, GDT, GDMP or BD1, see Scheme 1), and a 4-functional thiol crosslinker (PETMP) in the presence of a catalyst (TBD, typically 2.5 mol\%, unless otherwise mentioned). All molar fractions were measured with respect to the quantity of epoxy functions, which was taken as 100\%. The networks were accordingly labelled by the type of thiol spacer used. 

To prepare the starting solution, the epoxy monomer (BADGE, 100\% of epoxy functional groups, 1 g, 2.86 mmol), crosslinker (PETMP, 15\% of thiol functions, 0.110 g, 0.214 mmol), and GDMP (85\% of thiol functions, 2.43 mmol) were dissolved in anhydrous toluene (200 $\mu$L). Then, under vigorous stirring, the catalyst solution (TBD, 2.5 mol\% to the epoxy groups, 20 mg, 0.143 mmol, dissolved in 200 $\mu$L of anhydrous acetonitrile) was added. The mixture is poured into a PTFE mould and left to polymerise at ambient conditions for 30 min. After the polymerisation is completed, samples are transferred to a vacuum oven at 80$^\circ$C for 24 h to ensure complete solvent removal. The final material appears as a transparent elastomeric network. 

All materials are remoulded into a thin film after polymerisation using a hydraulic hot press, at a temperature 150-170$^\circ$C and pressure 1-2 MPa. All results reported are obtained using the  materials processed in this way (approximate thickness of samples: 0.7 mm). Samples were pressed for 20 min to ensure a sufficient bond exchange has taken place, and were left to cool overnight while still under pressure to yield a uniform transparent thin film. 

Variations of these compositions were also tested, altering the crosslinking density and changing the catalyst content (see Supplementary Information).

\textbf{Gel fraction procedure}. Samples of approximately 10 x 0.7 x 5 mm were immersed in a large volume (13-15 mL) of solvent for four days, with the solvent being renewed every day. Samples were then left to dry overnight in a covered glass culture dish at ambient conditions before being further dried in a vacuum oven. We tested several solvents and found that the best solubility (highest sol fraction) was in chloroform, while toluene swelled the networks less and correspondingly resulted in a lower sol fraction. 

However, one has to be somewhat cautious interpreting the results of swelling / gel-fraction experiments in vitrimers with transesterification BER.\cite{nicolay2019} As Fig. \ref{scheme2}(b) illustrates, there are going to be some fraction of OH-terminated chains, or loops, separated from the network altogether, once the bond exchange is initiated, and it has been separately demonstrated that swelling does initiate BER at low temperatures.\cite{yanji2018} So the sol fraction one obtains includes not only the unreacted products, but also those newly separated strands.

\textbf{Stress relaxation (``iso-strain'') measurements}. We used a home-made dynamic-mechanical instrument that allowed full control of the force applied to the sample, its deformation, and temperature. All our experiments were done in ambient air in an air-conditioned lab with 70\% humidity maintained. The heated sample chamber had a glass front end to allow optical-tracking of sample dimensions. After mounting, the samples were brought to the taut length and allowed to relax at the chosen temperature until full equilibrium was assured (by monitoring the tension force reaching a flat plateau). Then the step strain of 3-5\% was applied at a high rate (in less than 1 sec) and the tensile force $F(t)$ exerted by the stretched sample was monitored over a long time. At all times, all samples remained in rubber-elastic state, over a 100$^\circ$C above their glass transition.  At such a small extension, in the linear elastic regime, the sample cross-section was changing very little, so we were conducting the ``iso-strain'' test measuring the tensile stress. The raw data on relaxation of the force were collected, and processed to report the normalised relaxation function $F(t)/F_\mathrm{max}$, equivalent to the normalised stress $\sigma(t)/\sigma_\mathrm{max}$, which was presented in the plots.

\textbf{Constant-force (``iso-stress'') experiments}.  In the literature, these tests are often referred to as ``dilatometry'', which is a wrong term since the volume of the polymer network stays strictly constant. We used the TA DMA 850 instrument in tensile mode. The samples were initially equilibrated at a starting temperature of 70$^\circ$C, confirming that no relaxing tension remained, after which the stress of 50 kPa was applied. The resulting extensional strain was a reflection of the dynamic Young modulus of each network. The extensional strain of a sample under constant tensile stress was then monitored as they were subject to a ramp in temperature of 2$^\circ$C/min, as the material started to flow at a certain temperature. 

\textbf{Other characterisation}. Thermal stability was tested in thermal gravimetry analysis (TGA), using TA Instruments Q500 TGA, with a temperature ramp of 10$^\circ$C/min under argon atmosphere. Differential scanning calorimetry (DSC) analysis was done on a Perkin Elmer DSC 4000 instrument. A temperature ramp of 10$^\circ$C/min was used. Samples were subjected to a heating-cooling-heating cycle from -50 to 140$^\circ$C; data was extracted from the second heating run to eliminate the effects of thermal history. Infrared spectrometry (FTIR) data were recorded between 400-4000 cm$^{-1}$ on a Thermo Scientific Nicolet iS10 spectrometer using KBr Real Crystal IR sample cards and cover slips. For monitoring polymerisation conversion, the disappearance of the thiol peak was used as tracker of the advancement of the reaction.  

%%%%%%%%%%%%%%%%%%%%%%%%%%%%%%%%%%%%%%%%%
\section{Results and Discussion}

A facile epoxy-thiol polymerisation reaction was used in this work as a replacement of the traditional epoxy-acid reaction for the preparation of vitrimers.\cite{leibler2011,xlce2014} This change in chemistry results in drastically milder polymerisation conditions compared with epoxy-acid polymerisation, with the reaction generally completed at room temperature within one hour. The use of a thiol ``click'' reaction in these conditions results in the elimination of any undesired side reactions.\cite{bowman2010}  Additionally, epoxy homo-polymerisation of BADGE is inhibited by the low temperatures at which the reaction is conducted despite the presence of a base catalyst.\cite{fernandez2016} Similarly, the secondary amine catalyst employed (TBD), corresponding to 2.5 mol\% of the epoxy functional groups present (unless specified otherwise), is not expected to significantly bond into the network at such a temperature,\cite{wan2012} resulting in a minimal disruption of the network structure. Hence the resulting material presents an overall controlled and uniform network, with its crosslinking density being solely dictated by the concentration of crosslinking monomer within the initial reactive mixture. FTIR confirmed an essentially complete conversion of the monomers via monitoring of the consumption of the thiol species, as indicated in Supplementary Information. In the traditional epoxy-acid polymerisation method, the ester and hydroxyl groups that are required for transesterification are generated in-situ during the polymerisation; their concentration is hence globally fixed. However, in the case of epoxy-thiol chemistry, the ester groups are only incorporated into the network via the internal structure of the thiol spacers and the cross-linker (see Fig. \ref{scheme1}). Although hydroxyl groups are still generated \textit{in situ} during the polymerisation of epoxy groups, additional groups could be added through the structure of the thiol monomer GDT. This enables a broad control over the concentration of reactive functions for the BER within the network.  

Additionally, the use of a sulphur -S-, rather than oxygen -O-, in the backbone chain within the network results in more flexible chains, and, as a rule, a much lower glass transition. As a result, our materials  at ambient temperature remained elastomers, rather than glassy plastics as is common for epoxy-based networks.\cite{fernandez2016,lotti2006} The glass transition in our networks varied between 10$^\circ$C and 20$^\circ$C, details in the Supplementary Information. 

\begin{figure} 
\centering
\includegraphics[width=0.85\columnwidth]{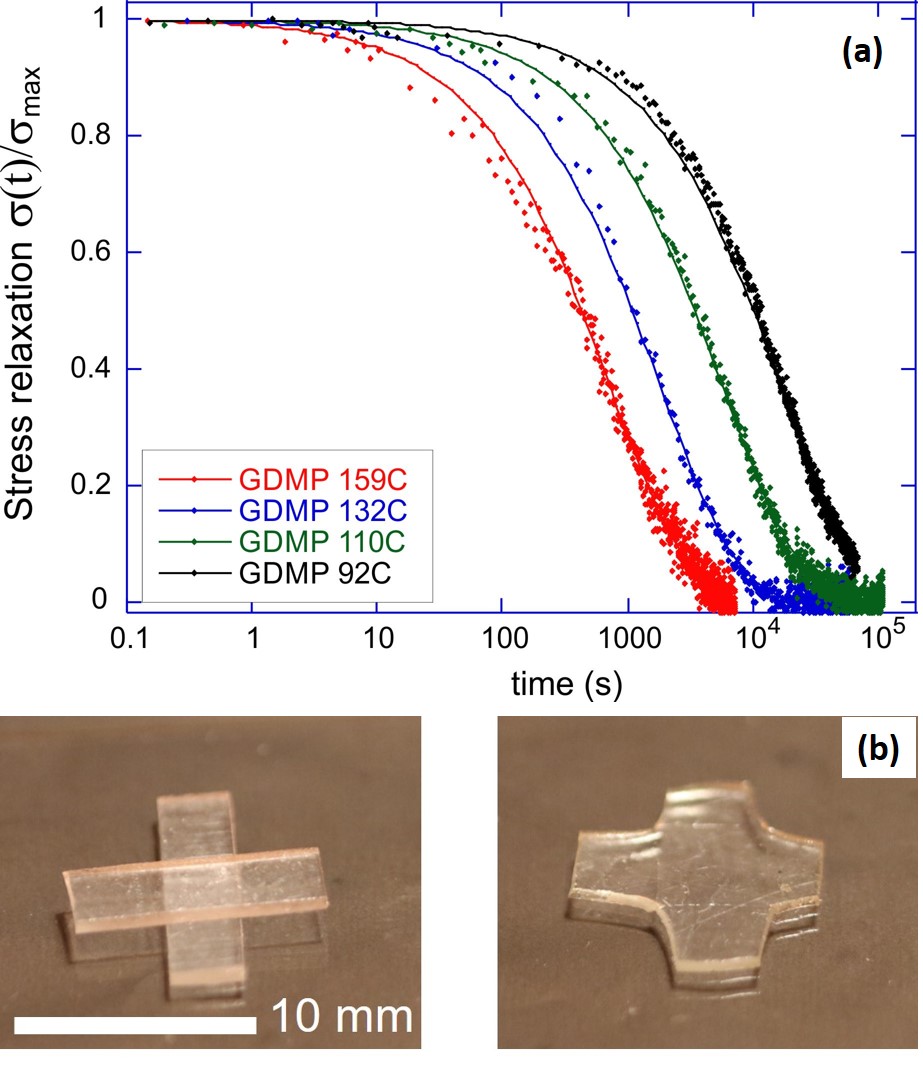}
\caption{a) An example of scaled stress relaxation curves for the GDMP vitrimer network with 2.5 mol\% of TBD catalyst and a 15\% crosslinking density, at several temperatures labelled in the plot. Good accuracy was obtained with simple exponential fitting. b) Two strips of GDMP elastomer were merged via hot pressing with a hydraulic press (160$^\circ$C and 40 MPa) to yield a single uniform material in the form of a cross, with no visible degradation. }
\label{fig1}
\end{figure}

Figure \ref{fig1}(a) shows an example of our typical stress-relaxation experiment, in this case using the GDMP network at several temperatures to illustrate how the characteristic relaxation time varies with temperature. The scaled relaxation curves were well fitted by a simple exponential function $exp[-t/\tau]$ with the relaxation time $\tau(T)$ displaying a temperature dependence following the Arrhenius-type activation law, which we analyse below. This behaviour is highly unusual in permanent polymer networks, as well as in regular thermoplastic elastomers, which also show the stress relaxation to zero, but never have the single relaxation time. The single relaxation time is a characteristic signature of the dynamic bond exchange in vitrimers.\cite{meng2016}

Experiments of sample welding through hot pressing further confirmed this result, demonstrating the complete re-moulding achieved in 5-10 min under stress of about 40 MPa at a temperature close to the elastic-plastic transition of a given network, Fig. \ref{fig1}(b).

\begin{figure} 
\centering
\includegraphics[width=0.99\columnwidth]{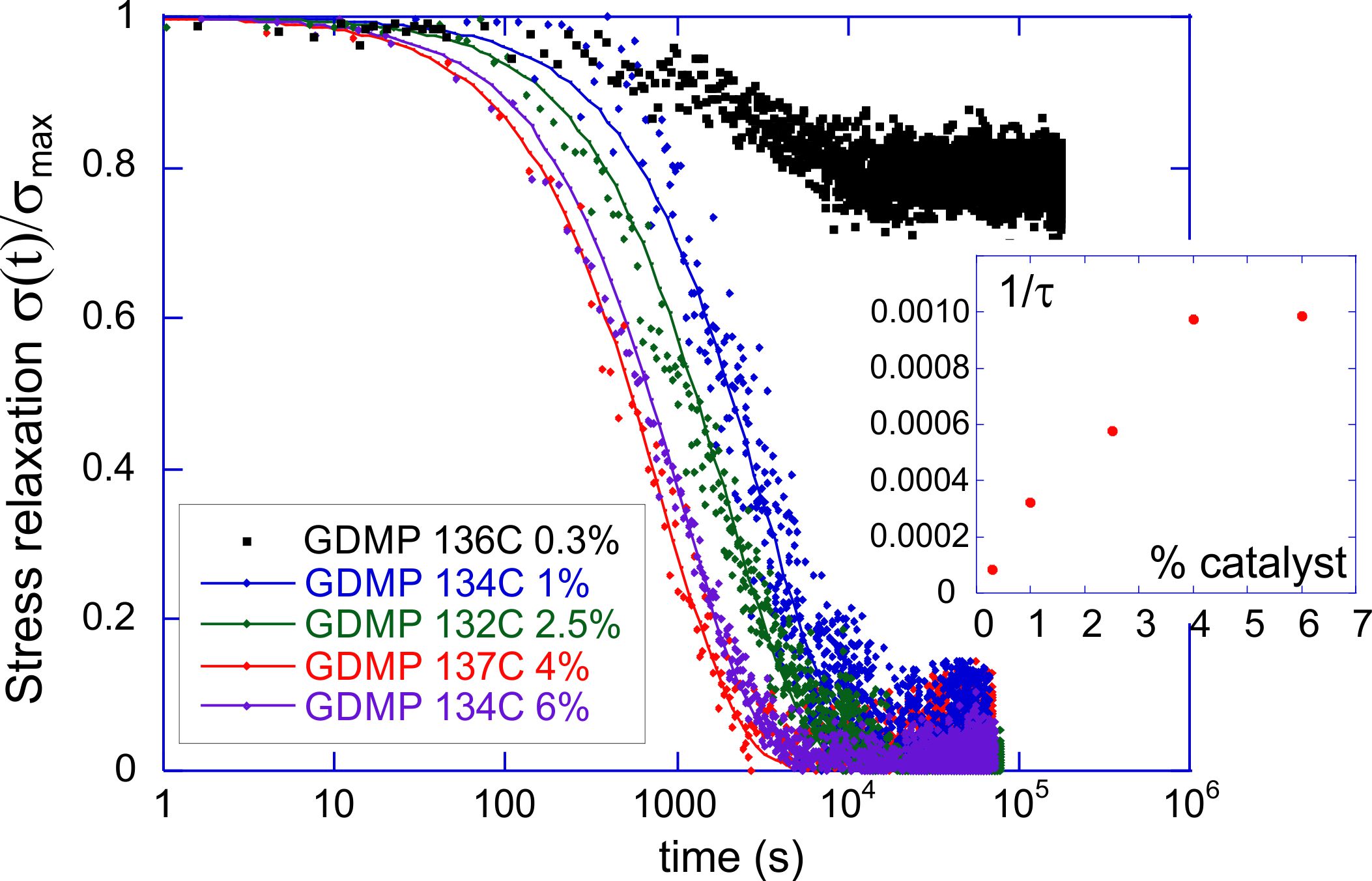}
\caption{Scaled stress-relaxation curves for GDMP vitrimers at T~135$^\circ$C and several TBD catalyst concentrations, as labelled in the plot. The inset shows the reaction rate: the inverse of relaxation time $\tau(T)$, in seconds, as a function of catalyst concentration, illustrating the saturation above 4 mol\%. }
\label{fig2}
\end{figure}

Here we investigate the influence of various factors on the kinetics of the elastic-plastic transition in transesterification-based vitrimers. Previous studies of such  systems demonstrated that the nature and the concentration of the catalyst play a key role in determining the dynamics of the BER.\cite{leibler2012cat}  A comparative study of the stress relaxation dynamics at different loadings of our catalyst (TBD) in a GDMP-based network demonstrate the expected increase in the speed of relaxation at a given temperature, which is synonymous to the lowering of the elastic-plastic transition temperature (Fig. \ref{fig2}). We however find here that this increase only carries until saturation beyond a certain catalyst loading (approximately 4-5 mol\%). This saturation can be understood as a competition between the kinetically limiting factors within the network: between the catalytic activity and the network chain mobility. As the bond exchange becomes easier due to an increase in catalytic loading, the limiting factor then becomes the physical encounter of matching reacting groups, leading to the saturation of relaxation time at a certain low value characteristic of the individual network. It could also be that the catalyst reaches a limit of its solubility in the network. 

Exploiting the advantages of epoxy-thiol chemistry, four different networks with a range of properties in terms of mechanical stiffness, activation energy of the BER, and concentration of reactive bonds, were synthesised. To enable a fair comparison, the crosslink density and the catalyst loading were kept constant across all four networks (15\% crosslink density and a 2.5 mol\% loading of TBD, respectively). 

An important parameter is the overall concentration of reactive functions required for transesterification within the networks. This parameter, labelled $N_\mathrm{func}$ here, was defined as the amount of reactive functions (both esters and hydroxyls) per single molecule of BADGE. As an example, a network with no ester functions would give  $N_\mathrm{func} = 2$, because two of the BADGE epoxy functions will produce a hydroxyl each. A network with exactly as many ester functions as hydroxyl functions would result in  $N_\mathrm{func} = 4$. 

For the GDMP network, there is a 1:1 ratio between hydroxyl and ester functions in the network (giving  $N_\mathrm{func} = 4$). The BD1 network has exactly the same function ratio than GDMP (and so the same  $N_\mathrm{func}$), but with an additional methyl group in the structure resulting in a more sterically hindered ester on the one hand, while on the other hand leading to a lower elastic modulus of the network by reducing the packing ability of the chains. Comparing to these two thiol spacers, EDDT gives a deficit in ester functions in the network (the only esters present in the EDDT network come from the crosslinker PETMP, giving $N_\mathrm{func} = 2.3$). At the other end of the scale, the GDT spacer has an additional hydroxyl function within its structure, giving $N_\mathrm{func} = 4.85$. The four materials possessed similar glass transitions, all within the range of 10 – 20$^\circ$C. This gap of less than 10 degrees between the highest and the lowest value of $T_g$ was deemed sufficiently small to compare the networks without any bias to the results originating from the proximity of the glass transition. 

\begin{figure} 
\centering
\includegraphics[width=0.85\columnwidth]{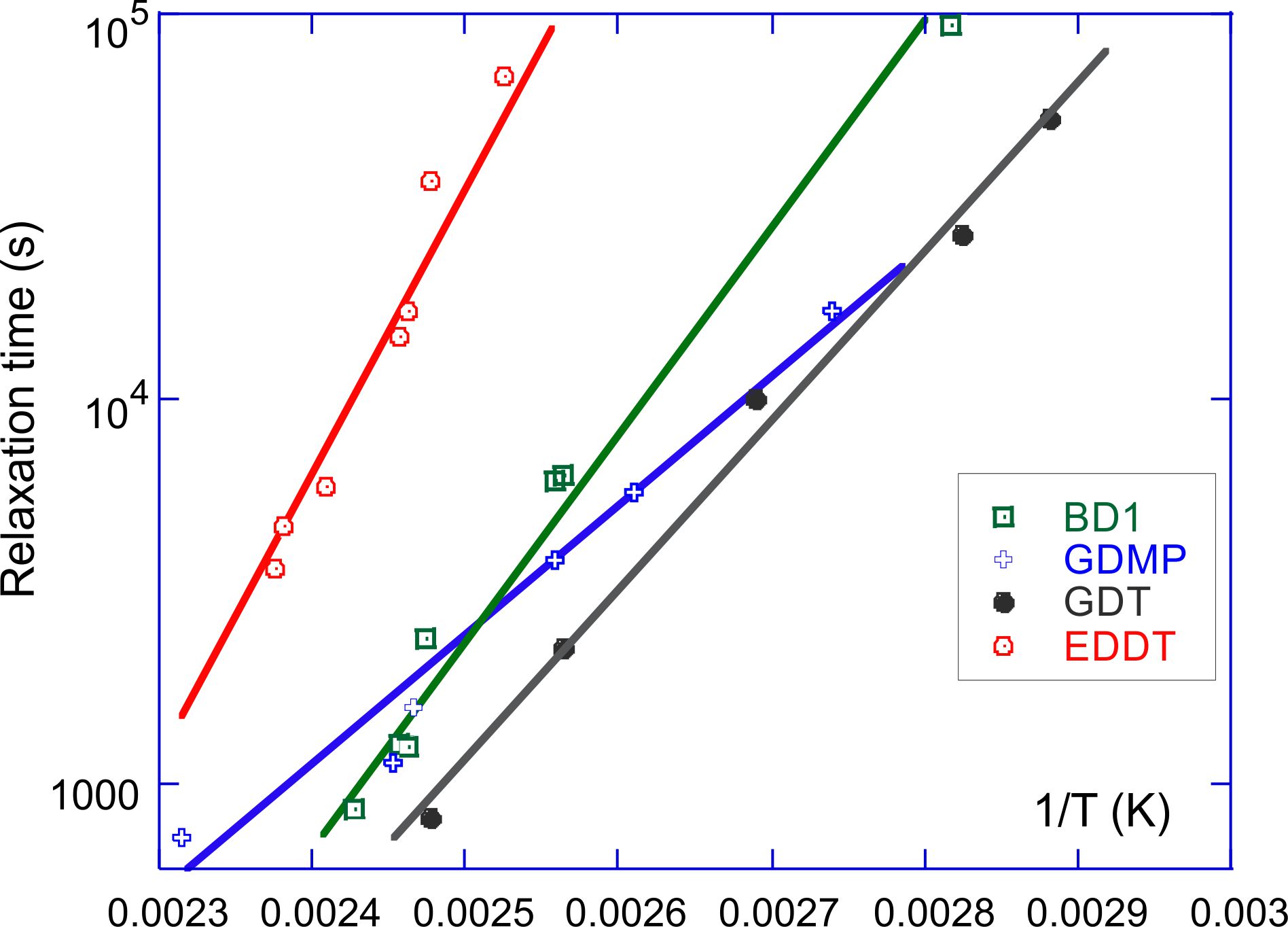}
\caption{The Arrhenius plots of the stress relaxation time $\tau(T)$ of our four different vitrimer networks (at 2.5 mol\% TBD and 15\% crosslinking density). The linear fitting of $\ln[\tau]$ vs. $1/T$ gives the activation energy $\Delta G$ of the bond exchange, which is listed in Table 1.}
\label{fig3}
\end{figure}

Figure \ref{fig3} shows the Arrhenius plot for all four materials. That is, we plot the logarithm of the relaxation time obtained in the stress-relaxation experiments, $\ln[\tau]=\mathrm{const}+\Delta G/k_BT$, taking the temperature in the absolute (Kelvin) units, and determine the slope of the resulting straight line. It is clear that all data sets show a clean single value of activation energy. We obtained the activation energy for transesterification: $\Delta G \approx$ 134 kJ/mol for EDDT, 92 kJ/mol for BD1, 75 kJ/mol for GDT, and 54 kJ/mol for GDMP. The higher activation energy of BD1, comparing with GDMP (both having the same $N_\mathrm{func}$), reflects the steric hindrance of the BD1 esters due to the presence of the neighbouring methyl groups. For reference, the highest value of $\Delta G$ for EDDT corresponds to about 60$k_BT$ at room temperature, which means transesterification is practically impossible. The lowest value of $\Delta G$ for GDMP corresponds to 24$k_BT$ at room temperature. In Leibler's case,\cite{leibler2012cat} the similar analysis gives $\Delta G \approx$ 100 kJ/mol (or 45$k_BT$ at room temperature) for the TBD catalyst at 5 mol\%, and $\Delta G \approx $ 83 kJ/mol (or 34$k_BT$) for the zinc acetate catalyst also at 5 mol\%.  It is also interesting to compare with the activation energy of the straight ester-ester bond exchange without any catalyst added, reported in the old papers, \cite{higgins1992,keller1993}  with values $\Delta G \approx$ 155 kJ/mol, evidently not very far from our EDDT result.

\begin{table}[h]
\centering
\begin{tabular}{r|ccc}
%\hline
 & $N_\mathrm{func}$ & E (MPa) & $\Delta G$ (kJ/mol) \\ \hline
EDDT  & 2.3 & 1.6 & 134 \\
BD1 & 4 & 0.32 & 92 \\
GDMP & 4 & 0.58 & 54 \\
GDT & 4.85 & 0.73 & 75 
\end{tabular}
\label{tab1}
\caption{Properties of the four vitrimer networks studied. The rubber modulus was measured at 70$^\circ$C: at least 50$^\circ$C above the glass transition of the materials to not be affected by the inter-sample variation of the $T_g$. }
\end{table}

\begin{figure} 
\centering
\includegraphics[width=0.85\columnwidth]{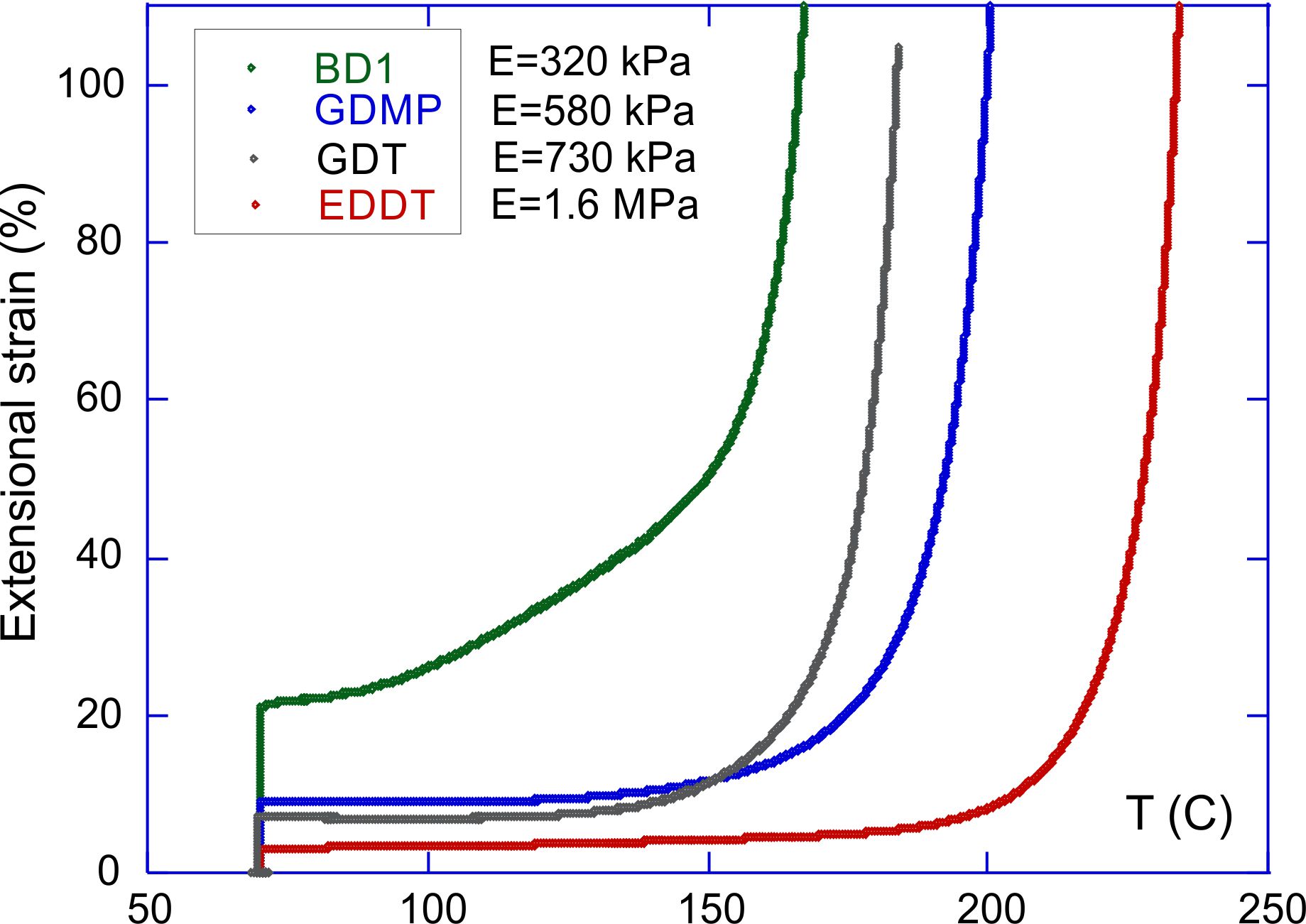}
\caption{The ‘iso-stress’ test: the sample extension at a constant engineering stress (50 kPa here), with the temperature increasing at 2$^\circ$C/min, comparing different network compositions, as labelled in the plot. All networks are with 15\% crosslinking density and 2.5\% TBD catalyst loading. The associated values of the Young modulus are also listed on the plot.}
\label{fig4}
\end{figure}

The elastic-plastic transition of these vitrimers was also characterised through the constant-force (iso-stress) experiments (Fig. \ref{fig4}). Surprisingly, the onset of plastic flow displayed by these materials did not follow the order of progression of the activation energy $\Delta G$ for the four materials studied, see Table 1. This observation invalidates the hypothesis that the activation energy has the dominant effect on the flow kinetics of vitrimers. Such a hypothesis would mean that the lowest temperature for the flow onset would be registered for the GDMP network (lowest  $\Delta G$), but instead the BD1 network starts its plastic flow first. Despite the more sterically hindered local environment in BD1, the iso-stress experiment shows that the dynamics of bond exchange proceeds more easily in the BD1 network compared to the GDMP network at any given temperature, suggesting that other factors have to be taken into account when predicting the macroscopic behaviour of vitrimers.

A suggestion becomes apparent when comparing the elastic-plastic transition in the iso-stress experiment with the rubber modulus of the networks (measured at 70$^\circ$C, well above the glass transition). The EDDT vitrimer has the highest activation energy (correlating with the lowest value for $N_\mathrm{func}$) and also the highest stiffness. It is consequently unsurprising that this material has the onset of flow at the highest temperature. 

BD1 displayed the lowest value for the Young modulus and the earliest onset of flow, despite having a mid-range value for both $N_\mathrm{func}$ and $\Delta G$. GDT and GDMP both show intermediate values for their modulus; however, despite GDT having a higher $\Delta G$ and a higher modulus, the onset of flow for GDT occurs at a lower temperature compared to GDMP.  

The onset of flow in the series GDT -- GDMP -- EDDT seems to follow the progression of $N_\mathrm{func}$ (and the activation energy $\Delta G$): a higher concentration of reactive groups leading to an earlier onset of the plastic flow on increasing the temperature. This stands to reason, as a higher concentration of reactive functions within the network enhances the probability of encounter of two reactive groups in the material under stress. In other words, the effective rate of attempts of BER increases $N_\mathrm{func}$. 

However, BD1 displays a behaviour contrary to this series: despite having the same $N_\mathrm{func}$ as GDMP (and so higher than GDT’s), and a $N_\mathrm{func}$ almost the double of GDMP’s, it still shows an onset of flow that is even earlier than GDT’s. The early onset of the flow in BD1 seems to be linked to the much lower stiffness of this material (Table 1). With that consideration in mind, the series BD1 -- GDMP -- EDDT seems to follow the order of increasing modulus for the flow onset. 

\begin{figure} 
\centering
\includegraphics[width=0.85\columnwidth]{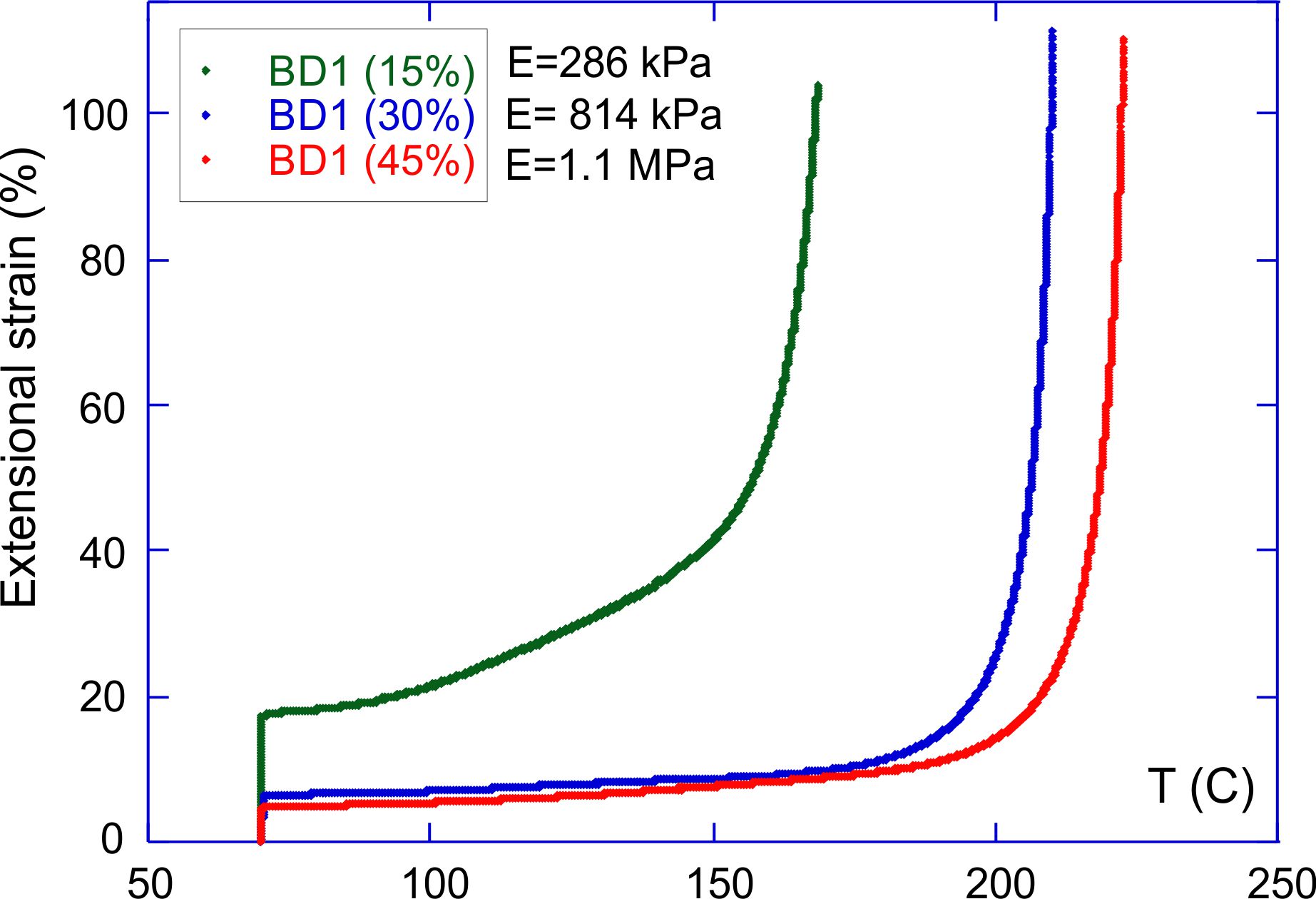}
\caption{The sample extension at a constant engineering stress of 50 kPa, with temperature increasing at 2$^\circ$C/min (the iso-stress test), comparing the same network composition with BD1 thiols, but increasing the crosslink density, as labelled in the plot. The associated values of the Young modulus are also listed on the plot, and we see a consistent shift of the apparent elastic-plastic transition temperature. }
\label{fig5}
\end{figure}

We therefore test the proposition that the mechanical stiffness of the rubbery network, as measured by the linear (Young) modulus E, is another factor defining the onset of the large plastic flow with increasing temperature (the catastrophic failure of the sample at constant stress). For this, we choose one of the vitrimers (BD1), and vary the crosslinking density in its network so that the stiffness of the elastomer changes in a controlled manner. At the same time, the activation energy of the BER within the networks is expected to be minimally affected by the change of crosslinking density, as $N_\mathrm{func}$ remains constant.\cite{hayashi2019a} Figure \ref{fig5} shows the result of the iso-stress experiment for BD1 vitrimers with the ‘standard’ 2.5 mol\% TBD, and the increasing 15\%, 30\%, and 45\% crosslinking density. A consistent increase of the elastic-to-plastic threshold was observed with the increase of the Young modulus of the network (Fig. \ref{fig5}). This confirms the key role played by the network elastic properties in defining the plastic-flow behaviour displayed by a vitrimer.

This can be rationalised by observing that in a rubbery network (given the glass transition in all our materials is below room temperature and the elastic-plastic transition occurs certainly above 150$^\circ$C), the local microscopic displacement that occurs after the chains re-connect after each instance of BER is approximately the mesh size. In other words, the longer the network strands are between the permanent crosslinks, the greater local deformation would be achieved when the event of transesterification occurs. So even at a slightly lower rate of the BER itself (due to the higher activation energy), the effective macroscopic flow could be higher in a network with a lower rubber modulus. 

A suggestion that the degree of crosslinking does have an effect on vitrimer relaxation has been recently put forward by Hayashi and Yano.\cite{hayashi2019b} In contrast to our observations, they found that   stress relaxation was faster in their more densely crosslinked networks. However, their network structure was different in the position and topology of hydroxyl groups between networks of different crosslink densities, and so the change in Young modulus in their networks was not the only result of changing composition. 

For samples with an identical network structure (i.e. at a given crosslinking density), the chain mobility, as measured through the elastic modulus, was hence shown to be critical in determining the threshold of plastic flow on the macroscopic scale. This factor is to be correlated with the ease of sample reprocessability. Indeed, sample remoulding through hot pressing with a hydraulic press followed the same temperature pattern and range as the one observed through iso-stress. 
In the literature, studies of vitrimer properties generally base their analysis solely on the stress relaxation response to draw conclusions.\cite{snyder2018,yu2014,hayashi2019b,brutman2019} Our stress relaxation experiments showed a characteristic trend across our samples (Fig. \ref{fig3}). At a given temperature (e.g. between 100 and 125$^\circ$C), the relaxation of the stress is the fastest for GDT, followed by GDMP, BD1, and finally EDDT. By extrapolating the linear fits on the Arrhenius graphs, it appears that GDT and BD1 should have similar relaxation times at 160$^\circ$C; this however contradicts the results observed in iso-stress (Fig. \ref{fig4}), where BD1 has a significantly higher flow rate than GDT at that temperature. The discrepancy between these two types of experiments is intriguing, and elucidating these differences would certainly provide a deeper understanding of vitrimer dynamics. 

We conclude that two competing and complementary factors dictate and control the plastic flow behaviour of vitrimers (for a given catalyst type and content): the concentration of reactive functions within the network and the elastic modulus of the material. The concentration and the local environment of the reactive functions affects the activation energy, and the rate of the bond exchange (transesterification). The rubber modulus of the network reflects the average mesh size, and through that – the magnitude of the local extension step that occurs on each event of the bond exchange.

\section{Conclusions}

In summary, an epoxy-thiol ``click'' chemistry was used for the polymerisation of transesterification-based vitrimers, as it enabled a fine control over the network structure. By varying the network spacers we were able to control the concentration of reactive functions within the network while guaranteeing a rigorously constant crosslinking density. This study revealed that the kinetics of the network flow in transesterification-based vitrimers was a result of the competition of two key factors: the elastic stiffness of the materials and the concentration and local environment of the hydroxyls and esters within the network. We believe that such findings offer a deeper understanding by raising questions relevant for all vitrimer systems, and suggesting answers that are supposed to be general. Further work is necessary to develop a more comprehensive theoretical model incorporating these findings to predict vitrimer properties based on the network characteristics. 

\section*{Acknowledgements}
We are grateful for the technical help of Herman Guo, and advice and discussions with Takuya Ohzono and Yan Ji. This work was supported by the ERC AdG ``APRA'' (786659).

\section*{References}

\providecommand*{\mcitethebibliography}{\thebibliography}
\csname @ifundefined\endcsname{endmcitethebibliography}
{\let\endmcitethebibliography\endthebibliography}{}


\begin{mcitethebibliography}{54}
\providecommand*{\natexlab}[1]{#1}
\providecommand*{\mciteSetBstSublistMode}[1]{}
\providecommand*{\mciteSetBstMaxWidthForm}[2]{}
\providecommand*{\mciteBstWouldAddEndPuncttrue}
  {\def\EndOfBibitem{\unskip.}}
\providecommand*{\mciteBstWouldAddEndPunctfalse}
  {\let\EndOfBibitem\relax}
\providecommand*{\mciteSetBstMidEndSepPunct}[3]{}
\providecommand*{\mciteSetBstSublistLabelBeginEnd}[3]{}
\providecommand*{\EndOfBibitem}{}
\mciteSetBstSublistMode{f}
\mciteSetBstMaxWidthForm{subitem}
{(\emph{\alph{mcitesubitemcount}})}
\mciteSetBstSublistLabelBeginEnd{\mcitemaxwidthsubitemform\space}
{\relax}{\relax}

\bibitem[Montarnal \emph{et~al.}(2011)Montarnal, Capelot, Tournilhac, and
  Leibler]{leibler2011}
D.~Montarnal, M.~Capelot, F.~Tournilhac and L.~Leibler, \emph{Science}, 2011,
  \textbf{334}, 965--968\relax
\mciteBstWouldAddEndPuncttrue
\mciteSetBstMidEndSepPunct{\mcitedefaultmidpunct}
{\mcitedefaultendpunct}{\mcitedefaultseppunct}\relax
\EndOfBibitem
\bibitem[Denissen \emph{et~al.}(2016)Denissen, Winne, and
  Du~Prez]{denissen2016}
W.~Denissen, J.~M. Winne and F.~E. Du~Prez, \emph{Chem. Sci.}, 2016,
  \textbf{7}, 30--38\relax
\mciteBstWouldAddEndPuncttrue
\mciteSetBstMidEndSepPunct{\mcitedefaultmidpunct}
{\mcitedefaultendpunct}{\mcitedefaultseppunct}\relax
\EndOfBibitem
\bibitem[Krishnakumar \emph{et~al.}(2020)Krishnakumar, Sanka, Binder,
  Parthasarthy, Rana, and Karak]{krishna2020}
B.~Krishnakumar, R.~P. Sanka, W.~H. Binder, V.~Parthasarthy, S.~Rana and
  N.~Karak, \emph{Chem. Eng. J.}, 2020, \textbf{385}, 123820\relax
\mciteBstWouldAddEndPuncttrue
\mciteSetBstMidEndSepPunct{\mcitedefaultmidpunct}
{\mcitedefaultendpunct}{\mcitedefaultseppunct}\relax
\EndOfBibitem
\bibitem[Cash \emph{et~al.}(2015)Cash, Kubo, Bapat, and Sumerlin]{cash2015}
J.~J. Cash, T.~Kubo, A.~P. Bapat and B.~S. Sumerlin, \emph{Macromolecules},
  2015, \textbf{48}, 2098--2106\relax
\mciteBstWouldAddEndPuncttrue
\mciteSetBstMidEndSepPunct{\mcitedefaultmidpunct}
{\mcitedefaultendpunct}{\mcitedefaultseppunct}\relax
\EndOfBibitem
\bibitem[Chen \emph{et~al.}(2019)Chen, Li, Wei, Venerus, and
  Torkelson]{chen2019}
X.~Chen, L.~Li, T.~Wei, D.~C. Venerus and J.~M. Torkelson, \emph{ACS Appl. Mat.
  Interfaces}, 2019, \textbf{11}, 2398--2407\relax
\mciteBstWouldAddEndPuncttrue
\mciteSetBstMidEndSepPunct{\mcitedefaultmidpunct}
{\mcitedefaultendpunct}{\mcitedefaultseppunct}\relax
\EndOfBibitem
\bibitem[Pei \emph{et~al.}(2014)Pei, Yang, Chen, Terentjev, Wei, and
  Ji]{xlce2014}
Z.~Pei, Y.~Yang, Q.~Chen, E.~M. Terentjev, Y.~Wei and Y.~Ji, \emph{Nature
  Mater.}, 2014, \textbf{13}, 36--41\relax
\mciteBstWouldAddEndPuncttrue
\mciteSetBstMidEndSepPunct{\mcitedefaultmidpunct}
{\mcitedefaultendpunct}{\mcitedefaultseppunct}\relax
\EndOfBibitem
\bibitem[Yang \emph{et~al.}(2018)Yang, Terentjev, Wei, and Ji]{yanji2018}
Y.~Yang, E.~M. Terentjev, Y.~Wei and Y.~Ji, \emph{Nature Comm.}, 2018,
  \textbf{9}, 1906\relax
\mciteBstWouldAddEndPuncttrue
\mciteSetBstMidEndSepPunct{\mcitedefaultmidpunct}
{\mcitedefaultendpunct}{\mcitedefaultseppunct}\relax
\EndOfBibitem
\bibitem[R\"ottger \emph{et~al.}(2017)R\"ottger, Domenech, van~der Weegen,
  Breuillac, Nicola\"y, and Leibler]{leibler2017bo}
M.~R\"ottger, T.~Domenech, R.~van~der Weegen, A.~Breuillac, R.~Nicola\"y and
  L.~Leibler, \emph{Science}, 2017, \textbf{356}, 62--65\relax
\mciteBstWouldAddEndPuncttrue
\mciteSetBstMidEndSepPunct{\mcitedefaultmidpunct}
{\mcitedefaultendpunct}{\mcitedefaultseppunct}\relax
\EndOfBibitem
\bibitem[Saed \emph{et~al.}(2019)Saed, Gablier, and Terentjev]{gablier2019}
M.~O. Saed, A.~Gablier and E.~M. Terentjev, \emph{Adv. Func. Mater.}, 2019,
  1906458\relax
\mciteBstWouldAddEndPuncttrue
\mciteSetBstMidEndSepPunct{\mcitedefaultmidpunct}
{\mcitedefaultendpunct}{\mcitedefaultseppunct}\relax
\EndOfBibitem
\bibitem[Cromwell \emph{et~al.}(2015)Cromwell, Chung, and Guan]{guan2015}
O.~R. Cromwell, J.~Chung and Z.~Guan, \emph{J. Am. Chem. Soc.}, 2015,
  \textbf{137}, 6492--6495\relax
\mciteBstWouldAddEndPuncttrue
\mciteSetBstMidEndSepPunct{\mcitedefaultmidpunct}
{\mcitedefaultendpunct}{\mcitedefaultseppunct}\relax
\EndOfBibitem
\bibitem[Lu \emph{et~al.}(2012)Lu, Tournilhac, Leibler, and Guan]{guan2012}
Y.-X. Lu, F.~Tournilhac, L.~Leibler and Z.~Guan, \emph{J. Amer. Chem. Soc.},
  2012, \textbf{134}, 8424--8427\relax
\mciteBstWouldAddEndPuncttrue
\mciteSetBstMidEndSepPunct{\mcitedefaultmidpunct}
{\mcitedefaultendpunct}{\mcitedefaultseppunct}\relax
\EndOfBibitem
\bibitem[Denissen \emph{et~al.}(2015)Denissen, Rivero, NicolaÃ¿, Leibler,
  Winne, and Du~Prez]{denissen2015}
W.~Denissen, G.~Rivero, R.~NicolaÃ¿, L.~Leibler, J.~M. Winne and F.~E. Du~Prez,
  \emph{Adv. Fun. Mater.}, 2015, \textbf{25}, 2451--2457\relax
\mciteBstWouldAddEndPuncttrue
\mciteSetBstMidEndSepPunct{\mcitedefaultmidpunct}
{\mcitedefaultendpunct}{\mcitedefaultseppunct}\relax
\EndOfBibitem
\bibitem[Denissen \emph{et~al.}(2017)Denissen, Droesbeke, NicolaÃ¿, Leibler,
  Winne, and Du~Prez]{denissen2017}
W.~Denissen, M.~Droesbeke, R.~NicolaÃ¿, L.~Leibler, J.~M. Winne and F.~E.
  Du~Prez, \emph{Nature Comm.}, 2017, \textbf{8}, 14857\relax
\mciteBstWouldAddEndPuncttrue
\mciteSetBstMidEndSepPunct{\mcitedefaultmidpunct}
{\mcitedefaultendpunct}{\mcitedefaultseppunct}\relax
\EndOfBibitem
\bibitem[Guerre \emph{et~al.}(2018)Guerre, Taplan, Nicola\"y, Winne, and
  Du~Prez]{guerre2018}
M.~Guerre, C.~Taplan, R.~Nicola\"y, J.~M. Winne and F.~E. Du~Prez, \emph{J. Am.
  Chem. Soc.}, 2018, \textbf{140}, 13272--13284\relax
\mciteBstWouldAddEndPuncttrue
\mciteSetBstMidEndSepPunct{\mcitedefaultmidpunct}
{\mcitedefaultendpunct}{\mcitedefaultseppunct}\relax
\EndOfBibitem
\bibitem[Stukenbroeker \emph{et~al.}(2017)Stukenbroeker, Wang, Winne, Du~Prez,
  Nicola\"y, and Leibler]{duprez2017}
T.~Stukenbroeker, W.~Wang, J.~M. Winne, F.~E. Du~Prez, R.~Nicola\"y and
  L.~Leibler, \emph{Polymer Chem.}, 2017, \textbf{8}, 6590--6593\relax
\mciteBstWouldAddEndPuncttrue
\mciteSetBstMidEndSepPunct{\mcitedefaultmidpunct}
{\mcitedefaultendpunct}{\mcitedefaultseppunct}\relax
\EndOfBibitem
\bibitem[Nishimura \emph{et~al.}(2017)Nishimura, Chung, Muradyan, and
  Guan]{guan2017}
Y.~Nishimura, J.~Chung, H.~Muradyan and Z.~Guan, \emph{J. Amer. Chem. Soc.},
  2017, \textbf{139}, 14881--14884\relax
\mciteBstWouldAddEndPuncttrue
\mciteSetBstMidEndSepPunct{\mcitedefaultmidpunct}
{\mcitedefaultendpunct}{\mcitedefaultseppunct}\relax
\EndOfBibitem
\bibitem[Fortman \emph{et~al.}(2015)Fortman, Brutman, Cramer, Hillmyer, and
  Dichtel]{fortman2015}
D.~J. Fortman, J.~P. Brutman, C.~J. Cramer, M.~A. Hillmyer and W.~R. Dichtel,
  \emph{J. Amer. Chem. Soc.}, 2015, \textbf{137}, 14019--14022\relax
\mciteBstWouldAddEndPuncttrue
\mciteSetBstMidEndSepPunct{\mcitedefaultmidpunct}
{\mcitedefaultendpunct}{\mcitedefaultseppunct}\relax
\EndOfBibitem
\bibitem[Ishibashi and Kalow(2018)]{ishibashi2018}
J.~S.~A. Ishibashi and J.~A. Kalow, \emph{ACS Macro Lett.}, 2018, \textbf{7},
  482--486\relax
\mciteBstWouldAddEndPuncttrue
\mciteSetBstMidEndSepPunct{\mcitedefaultmidpunct}
{\mcitedefaultendpunct}{\mcitedefaultseppunct}\relax
\EndOfBibitem
\bibitem[Hendriks \emph{et~al.}(2017)Hendriks, Waelkens, Winne, and
  Du~Prez]{hendriks2017}
B.~Hendriks, J.~Waelkens, J.~M. Winne and F.~E. Du~Prez, \emph{ACS Macro
  Lett.}, 2017, \textbf{6}, 930--934\relax
\mciteBstWouldAddEndPuncttrue
\mciteSetBstMidEndSepPunct{\mcitedefaultmidpunct}
{\mcitedefaultendpunct}{\mcitedefaultseppunct}\relax
\EndOfBibitem
\bibitem[Obadia \emph{et~al.}(2015)Obadia, Mudraboyina, Serghei, Montarnal, and
  Drockenmuller]{obadia2015}
M.~M. Obadia, B.~P. Mudraboyina, A.~Serghei, D.~Montarnal and E.~Drockenmuller,
  \emph{J. Am. Chem. Soc.}, 2015, \textbf{137}, 6078--6083\relax
\mciteBstWouldAddEndPuncttrue
\mciteSetBstMidEndSepPunct{\mcitedefaultmidpunct}
{\mcitedefaultendpunct}{\mcitedefaultseppunct}\relax
\EndOfBibitem
\bibitem[Tang \emph{et~al.}(2017)Tang, Liu, Guo, and Zhang]{tang2017}
Z.~Tang, Y.~Liu, B.~Guo and L.~Zhang, \emph{Macromolecules}, 2017, \textbf{50},
  7584--7592\relax
\mciteBstWouldAddEndPuncttrue
\mciteSetBstMidEndSepPunct{\mcitedefaultmidpunct}
{\mcitedefaultendpunct}{\mcitedefaultseppunct}\relax
\EndOfBibitem
\bibitem[Capelot \emph{et~al.}(2012)Capelot, Unterlass, Tournilhac, and
  Leibler]{leibler2012cat}
M.~Capelot, M.~M. Unterlass, F.~Tournilhac and L.~Leibler, \emph{ACS Macro
  Lett.}, 2012, \textbf{1}, 789--792\relax
\mciteBstWouldAddEndPuncttrue
\mciteSetBstMidEndSepPunct{\mcitedefaultmidpunct}
{\mcitedefaultendpunct}{\mcitedefaultseppunct}\relax
\EndOfBibitem
\bibitem[Capelot \emph{et~al.}(2012)Capelot, Montarnal, Tournilhac, and
  Leibler]{leibler2012met}
M.~Capelot, D.~Montarnal, F.~Tournilhac and L.~Leibler, \emph{J. Am. Chem.
  Soc.}, 2012, \textbf{134}, 7664--7667\relax
\mciteBstWouldAddEndPuncttrue
\mciteSetBstMidEndSepPunct{\mcitedefaultmidpunct}
{\mcitedefaultendpunct}{\mcitedefaultseppunct}\relax
\EndOfBibitem
\bibitem[Yang \emph{et~al.}(2014)Yang, Pei, Zhang, Tao, Wei, and Ji]{yanji2014}
Y.~Yang, Z.~Pei, X.~Zhang, L.~Tao, Y.~Wei and Y.~Ji, \emph{Chem. Sci.}, 2014,
  \textbf{5}, 3486--3492\relax
\mciteBstWouldAddEndPuncttrue
\mciteSetBstMidEndSepPunct{\mcitedefaultmidpunct}
{\mcitedefaultendpunct}{\mcitedefaultseppunct}\relax
\EndOfBibitem
\bibitem[Pei \emph{et~al.}(2016)Pei, Yang, Chen, Wei, and Ji]{yanji2016}
Z.~Pei, Y.~Yang, Q.~Chen, Y.~Wei and Y.~Ji, \emph{Adv. Mater.}, 2016,
  \textbf{28}, 156--160\relax
\mciteBstWouldAddEndPuncttrue
\mciteSetBstMidEndSepPunct{\mcitedefaultmidpunct}
{\mcitedefaultendpunct}{\mcitedefaultseppunct}\relax
\EndOfBibitem
\bibitem[Zhao and Abu-Omar(2019)]{zhao2019}
S.~Zhao and M.~M. Abu-Omar, \emph{Macromolecules}, 2019, \textbf{52},
  3646--3654\relax
\mciteBstWouldAddEndPuncttrue
\mciteSetBstMidEndSepPunct{\mcitedefaultmidpunct}
{\mcitedefaultendpunct}{\mcitedefaultseppunct}\relax
\EndOfBibitem
\bibitem[Brutman \emph{et~al.}(2014)Brutman, Delgado, and
  Hillmyer]{Hillmyer2014}
J.~P. Brutman, P.~A. Delgado and M.~A. Hillmyer, \emph{ACS Macro Lett.}, 2014,
  \textbf{3}, 607--610\relax
\mciteBstWouldAddEndPuncttrue
\mciteSetBstMidEndSepPunct{\mcitedefaultmidpunct}
{\mcitedefaultendpunct}{\mcitedefaultseppunct}\relax
\EndOfBibitem
\bibitem[Chen \emph{et~al.}(2017)Chen, Yu, Pei, Yang, Wei, and Ji]{chen2017}
Q.~Chen, X.~Yu, Z.~Pei, Y.~Yang, Y.~Wei and Y.~Ji, \emph{Chem. Sci.}, 2017,
  \textbf{8}, 724--733\relax
\mciteBstWouldAddEndPuncttrue
\mciteSetBstMidEndSepPunct{\mcitedefaultmidpunct}
{\mcitedefaultendpunct}{\mcitedefaultseppunct}\relax
\EndOfBibitem
\bibitem[Hayashi \emph{et~al.}(2019)Hayashi, Yano, and Takasu]{hayashi2019a}
M.~Hayashi, R.~Yano and A.~Takasu, \emph{Polym. Chem}, 2019, \textbf{10},
  2047--2056\relax
\mciteBstWouldAddEndPuncttrue
\mciteSetBstMidEndSepPunct{\mcitedefaultmidpunct}
{\mcitedefaultendpunct}{\mcitedefaultseppunct}\relax
\EndOfBibitem
\bibitem[Altuna \emph{et~al.}(2019)Altuna, Hoppe, and Williams]{epoxy2019}
F.~I. Altuna, C.~E. Hoppe and R.~J. Williams, \emph{Euro. Polym. J.}, 2019,
  \textbf{113}, 297--304\relax
\mciteBstWouldAddEndPuncttrue
\mciteSetBstMidEndSepPunct{\mcitedefaultmidpunct}
{\mcitedefaultendpunct}{\mcitedefaultseppunct}\relax
\EndOfBibitem
\bibitem[Shi \emph{et~al.}(2017)Shi, Yu, Kuang, Mu, Dunn, Dunn, Wang, and
  Jerry~Qi]{shi2017}
Q.~Shi, K.~Yu, X.~Kuang, X.~Mu, C.~K. Dunn, M.~L. Dunn, T.~Wang and
  H.~Jerry~Qi, \emph{Mater. Horizons}, 2017, \textbf{4}, 598--607\relax
\mciteBstWouldAddEndPuncttrue
\mciteSetBstMidEndSepPunct{\mcitedefaultmidpunct}
{\mcitedefaultendpunct}{\mcitedefaultseppunct}\relax
\EndOfBibitem
\bibitem[Han \emph{et~al.}(2018)Han, Liu, Hao, Zhang, Guo, and
  Zhang]{catalyst-free2018}
J.~Han, T.~Liu, C.~Hao, S.~Zhang, B.~Guo and J.~Zhang, \emph{Macromolecules},
  2018, \textbf{51}, 6789--6799\relax
\mciteBstWouldAddEndPuncttrue
\mciteSetBstMidEndSepPunct{\mcitedefaultmidpunct}
{\mcitedefaultendpunct}{\mcitedefaultseppunct}\relax
\EndOfBibitem
\bibitem[Mat\v{e}jka \emph{et~al.}(1982)Mat\v{e}jka, Pokom\'{y}, and
  Du\v{s}ek]{matejka1982}
L.~Mat\v{e}jka, S.~Pokom\'{y} and K.~Du\v{s}ek, \emph{Polymer Bull.}, 1982,
  \textbf{7}, 123--128\relax
\mciteBstWouldAddEndPuncttrue
\mciteSetBstMidEndSepPunct{\mcitedefaultmidpunct}
{\mcitedefaultendpunct}{\mcitedefaultseppunct}\relax
\EndOfBibitem
\bibitem[Li \emph{et~al.}(2018)Li, Chen, Jin, and Torkelson]{li2018}
L.~Li, X.~Chen, K.~Jin and J.~M. Torkelson, \emph{Macromolecules}, 2018,
  \textbf{51}, 5537--5546\relax
\mciteBstWouldAddEndPuncttrue
\mciteSetBstMidEndSepPunct{\mcitedefaultmidpunct}
{\mcitedefaultendpunct}{\mcitedefaultseppunct}\relax
\EndOfBibitem
\bibitem[Meng \emph{et~al.}(2016)Meng, Pritchard, and Terentjev]{meng2016}
F.~Meng, R.~H. Pritchard and E.~M. Terentjev, \emph{Macromolecules}, 2016,
  \textbf{49}, 2843--2852\relax
\mciteBstWouldAddEndPuncttrue
\mciteSetBstMidEndSepPunct{\mcitedefaultmidpunct}
{\mcitedefaultendpunct}{\mcitedefaultseppunct}\relax
\EndOfBibitem
\bibitem[Meng \emph{et~al.}(2019)Meng, Saed, and Terentjev]{meng2019}
F.~Meng, M.~O. Saed and E.~M. Terentjev, \emph{Macromolecules}, 2019,
  \textbf{52}, 7423--7429\relax
\mciteBstWouldAddEndPuncttrue
\mciteSetBstMidEndSepPunct{\mcitedefaultmidpunct}
{\mcitedefaultendpunct}{\mcitedefaultseppunct}\relax
\EndOfBibitem
\bibitem[Self \emph{et~al.}(2018)Self, Dolinski, Zayas, Read~de Alaniz, and
  Bates]{self2018}
J.~L. Self, N.~D. Dolinski, M.~S. Zayas, J.~Read~de Alaniz and C.~M. Bates,
  \emph{ACS Macro Lett.}, 2018, \textbf{7}, 817--821\relax
\mciteBstWouldAddEndPuncttrue
\mciteSetBstMidEndSepPunct{\mcitedefaultmidpunct}
{\mcitedefaultendpunct}{\mcitedefaultseppunct}\relax
\EndOfBibitem
\bibitem[Snyder \emph{et~al.}(2018)Snyder, Fortman, De~Hoe, Hillmyer, and
  Dichtel]{snyder2018}
R.~L. Snyder, D.~J. Fortman, G.~X. De~Hoe, M.~A. Hillmyer and W.~R. Dichtel,
  \emph{Macromolecules}, 2018, \textbf{51}, 389--397\relax
\mciteBstWouldAddEndPuncttrue
\mciteSetBstMidEndSepPunct{\mcitedefaultmidpunct}
{\mcitedefaultendpunct}{\mcitedefaultseppunct}\relax
\EndOfBibitem
\bibitem[Yu \emph{et~al.}(2014)Yu, Taynton, Zhang, Dunn, and Qi]{yu2014}
K.~Yu, P.~Taynton, W.~Zhang, M.~L. Dunn and H.~J. Qi, \emph{RSC Adv.}, 2014,
  \textbf{4}, 48682--48690\relax
\mciteBstWouldAddEndPuncttrue
\mciteSetBstMidEndSepPunct{\mcitedefaultmidpunct}
{\mcitedefaultendpunct}{\mcitedefaultseppunct}\relax
\EndOfBibitem
\bibitem[Hayashi and Yano(2019)]{hayashi2019b}
M.~Hayashi and R.~Yano, \emph{Macromolecules}, 2019, \textbf{53},
  182--189\relax
\mciteBstWouldAddEndPuncttrue
\mciteSetBstMidEndSepPunct{\mcitedefaultmidpunct}
{\mcitedefaultendpunct}{\mcitedefaultseppunct}\relax
\EndOfBibitem
\bibitem[Brutman \emph{et~al.}(2019)Brutman, Fortman, De~Hoe, Dichtel, and
  Hillmyer]{brutman2019}
J.~P. Brutman, D.~J. Fortman, G.~X. De~Hoe, W.~R. Dichtel and M.~A. Hillmyer,
  \emph{J. Phys. Chem. B}, 2019, \textbf{123}, 1432--1441\relax
\mciteBstWouldAddEndPuncttrue
\mciteSetBstMidEndSepPunct{\mcitedefaultmidpunct}
{\mcitedefaultendpunct}{\mcitedefaultseppunct}\relax
\EndOfBibitem
\bibitem[Hoyle \emph{et~al.}(2010)Hoyle, Lowe, and Bowman]{bowman2010}
C.~E. Hoyle, A.~B. Lowe and C.~N. Bowman, \emph{Chem. Soc. Rev.}, 2010,
  \textbf{39}, 1355\relax
\mciteBstWouldAddEndPuncttrue
\mciteSetBstMidEndSepPunct{\mcitedefaultmidpunct}
{\mcitedefaultendpunct}{\mcitedefaultseppunct}\relax
\EndOfBibitem
\bibitem[Br\"andle and Khan(2012)]{brandle2012}
A.~Br\"andle and A.~Khan, \emph{Polymer Chemistry}, 2012, \textbf{3},
  3224\relax
\mciteBstWouldAddEndPuncttrue
\mciteSetBstMidEndSepPunct{\mcitedefaultmidpunct}
{\mcitedefaultendpunct}{\mcitedefaultseppunct}\relax
\EndOfBibitem
\bibitem[{Le Neindre} and Nicola\"y(2014)]{nicolay2014}
M.~{Le Neindre} and R.~Nicola\"y, \emph{Polymer Chem.}, 2014, \textbf{5},
  4601--4611\relax
\mciteBstWouldAddEndPuncttrue
\mciteSetBstMidEndSepPunct{\mcitedefaultmidpunct}
{\mcitedefaultendpunct}{\mcitedefaultseppunct}\relax
\EndOfBibitem
\bibitem[Belmonte \emph{et~al.}(2017)Belmonte, Russo, Ambrogi,
  Fern\'andez-Francos, and {De la Flor}]{belmonte2017}
A.~Belmonte, C.~Russo, V.~Ambrogi, X.~Fern\'andez-Francos and S.~{De la Flor},
  \emph{Polymers}, 2017, \textbf{9}, 113\relax
\mciteBstWouldAddEndPuncttrue
\mciteSetBstMidEndSepPunct{\mcitedefaultmidpunct}
{\mcitedefaultendpunct}{\mcitedefaultseppunct}\relax
\EndOfBibitem
\bibitem[Jin \emph{et~al.}(2016)Jin, Wilmot, Heath, and Torkelson]{jin2016}
K.~Jin, N.~Wilmot, W.~H. Heath and J.~M. Torkelson, \emph{Macromolecules},
  2016, \textbf{49}, 4115--4123\relax
\mciteBstWouldAddEndPuncttrue
\mciteSetBstMidEndSepPunct{\mcitedefaultmidpunct}
{\mcitedefaultendpunct}{\mcitedefaultseppunct}\relax
\EndOfBibitem
\bibitem[Fern\'andez-Francos \emph{et~al.}(2016)Fern\'andez-Francos, Konuray,
  Belmonte, De~la Flor, Serra, and Ramis]{fernandez2016}
X.~Fern\'andez-Francos, A.-O. Konuray, A.~Belmonte, S.~De~la Flor, A.~Serra and
  X.~Ramis, \emph{Polym. Chem.}, 2016, \textbf{7}, 2280--2290\relax
\mciteBstWouldAddEndPuncttrue
\mciteSetBstMidEndSepPunct{\mcitedefaultmidpunct}
{\mcitedefaultendpunct}{\mcitedefaultseppunct}\relax
\EndOfBibitem
\bibitem[Stuparu and Khan(2016)]{khan2016}
M.~C. Stuparu and A.~Khan, \emph{J. Polym. Sci. Part A: Polym. Chem.}, 2016,
  \textbf{54}, 3057--3070\relax
\mciteBstWouldAddEndPuncttrue
\mciteSetBstMidEndSepPunct{\mcitedefaultmidpunct}
{\mcitedefaultendpunct}{\mcitedefaultseppunct}\relax
\EndOfBibitem
\bibitem[Konuray \emph{et~al.}(2017)Konuray, Fern\'andez-Francos, and
  Ramis]{konuray2017}
A.~O. Konuray, X.~Fern\'andez-Francos and X.~Ramis, \emph{Polymer}, 2017,
  \textbf{116}, 191--203\relax
\mciteBstWouldAddEndPuncttrue
\mciteSetBstMidEndSepPunct{\mcitedefaultmidpunct}
{\mcitedefaultendpunct}{\mcitedefaultseppunct}\relax
\EndOfBibitem
\bibitem[Breuillac \emph{et~al.}(2019)Breuillac, Kassalias, and
  Nicola\"y]{nicolay2019}
A.~Breuillac, A.~Kassalias and R.~Nicola\"y, \emph{Macromolecules}, 2019,
  \textbf{52}, 7102--7113\relax
\mciteBstWouldAddEndPuncttrue
\mciteSetBstMidEndSepPunct{\mcitedefaultmidpunct}
{\mcitedefaultendpunct}{\mcitedefaultseppunct}\relax
\EndOfBibitem
\bibitem[Wan \emph{et~al.}(2012)Wan, Li, Bu, Xu, Li, and Fan]{wan2012}
J.~Wan, C.~Li, Z.-Y. Bu, C.-J. Xu, B.-G. Li and H.~Fan, \emph{Chem. Eng. J.},
  2012, \textbf{188}, 160--172\relax
\mciteBstWouldAddEndPuncttrue
\mciteSetBstMidEndSepPunct{\mcitedefaultmidpunct}
{\mcitedefaultendpunct}{\mcitedefaultseppunct}\relax
\EndOfBibitem
\bibitem[Lotti \emph{et~al.}(2006)Lotti, Siracusa, Finelli, Marchese, and
  Munari]{lotti2006}
N.~Lotti, V.~Siracusa, L.~Finelli, P.~Marchese and A.~Munari, \emph{Euro.
  Polym. J.}, 2006, \textbf{42}, 3374--3382\relax
\mciteBstWouldAddEndPuncttrue
\mciteSetBstMidEndSepPunct{\mcitedefaultmidpunct}
{\mcitedefaultendpunct}{\mcitedefaultseppunct}\relax
\EndOfBibitem
\bibitem[Arrighi \emph{et~al.}(1992)Arrighi, Higgins, Weiss, and
  Cimecioglu]{higgins1992}
V.~Arrighi, J.~S. Higgins, R.~A. Weiss and A.~L. Cimecioglu,
  \emph{Macromolecules}, 1992, \textbf{25}, 5297--5305\relax
\mciteBstWouldAddEndPuncttrue
\mciteSetBstMidEndSepPunct{\mcitedefaultmidpunct}
{\mcitedefaultendpunct}{\mcitedefaultseppunct}\relax
\EndOfBibitem
\bibitem[Li \emph{et~al.}(1993)Li, Br\'ulet, Keller, Strazielle, and
  Cotton]{keller1993}
M.~H. Li, A.~Br\'ulet, P.~Keller, C.~Strazielle and J.~P. Cotton,
  \emph{Macromolecules}, 1993, \textbf{26}, 119--124\relax
\mciteBstWouldAddEndPuncttrue
\mciteSetBstMidEndSepPunct{\mcitedefaultmidpunct}
{\mcitedefaultendpunct}{\mcitedefaultseppunct}\relax
\EndOfBibitem
\end{mcitethebibliography}
\end{document}